\documentclass{PoS}
\usepackage[]{gensymb}
\usepackage{enumerate}
\usepackage{siunitx}
\newcommand{\eh}[1]{\,\mathrm{#1}}
\newcommand{\gev}{\eh{GeV}}

\newcommand{\pcnt}{$\eh{\%}$}

\renewcommand{\epsilon}{\varepsilon}

\newcommand{\hess}{H.E.S.S.}
\newcommand{\nectar}{NECTAr}
\newcommand{\hessI}{H.E.S.S.~I}

\newcommand{\hessII}{H.E.S.S.~II}

\newcommand{\fref}[1]{Fig.~\ref{#1}}

\title{Performance of the upgraded H.E.S.S. cameras}

\ShortTitle{\hess\ upgrade performance}

\author{\speaker{S.~Bonnefoy} $^a$,
T.~Ashton$^b$,
M.~Backes$^h$,
A.~Balzer$^c$,
D.~Berge$^c$,
S.~Bonnefoy$^a$
F.~Brun$^d$,
T.~Chaminade$^d$,
E.~Delagnes$^d$,
G.~Fontaine$^f$,
M.~F\"u{\ss}ling$^a$,
G.~Giavitto$^a$,
B.~Giebels$^f$,
J.F.~Glicenstein$^d$,
T.~Gr\"aber$^a$,
J.A.~Hinton$^{b,g}$,
A.~Jahnke$^i$,
M.~Kossatz$^a$,
A.~Kretzschmann$^a$,
V.~Lefranc$^{a,d}$,
H.~Leich$^a$,
J.P.~Lenain$^e$,
H.~L\"udecke$^a$,
I.~Lypova$^a$,
P.~Manigot$^f$,
V.~Marandon$^g$,
E.~Moulin$^d$,
M.~de~Naurois$^f$,
P.~Nayman$^e$,
S.~Ohm$^a$,
M.~Penno$^a$,
D.~Ross$^b$,
D.~Salek$^c$,
M.~Schade$^a$,
T.~Schwab$^g$,
K.~Shiningayamwe$^h$
C.~Stegmann$^a$,
C.~Steppa$^a$,
J.~Thornhill$^b$,
F.~Toussenel$^e$\\
\llap{$^a$}DESY, D-15738 Zeuthen, Germany \\
\llap{$^b$}Department of Physics and Astronomy, The University of Leicester, University Road, Leicester, LE1 7RH, United Kingdom \\
\llap{$^c$}GRAPPA, Anton Pannekoek Institute for Astronomy, University of Amsterdam,  Science Park 904, 1098 XH Amsterdam, The Netherlands \\
\llap{$^d$}DSM/Irfu, CEA Saclay, F-91191 Gif-Sur-Yvette Cedex, France \\
\llap{$^e$}Sorbonne Universit\'es, UPMC Universit\'e Paris 06, Universit\'e Paris Diderot, Sorbonne Paris Cit\'e, CNRS, Laboratoire de Physique Nucl\'eaire et de Hautes Energies (LPNHE), 4 place Jussieu, F-75252, Paris Cedex 5, France \\
\llap{$^f$}Laboratoire Leprince-Ringuet, Ecole Polytechnique, CNRS/IN2P3, F-91128 Palaiseau, France \\
\llap{$^g$}Max-Planck-Institut f\"ur Kernphysik, P.O. Box 103980, D 69029 Heidelberg, Germany \\
\llap{$^h$}University of Namibia, Department of Physics, Private Bag 13301, Windhoek, Namibia\\
\llap{$^i$}JA consulting, St Michael Park 23, Avis, Windhoek, Namibia\\
}

\abstract{
The 14 years old cameras of the H.E.S.S. 12-m telescopes have been upgraded in
2015/2016, with the goals of reducing the system failure rate, reducing the
dead time and improving the overall performance of the array. This conference contribution
describes the various tests that were carried out on the cameras and their
sub-components both in the lab and on site. It also gives an overview of the
commissioning and calibration procedures adopted during and after the
installation, including e.g. flat-fielding and trigger threshold scans. Finally, it reports in detail about the overall performance of the four new \hessI \ cameras, using very recent data.
}

\FullConference{35th International Cosmic Ray Conference --- ICRC2017\\
		10--20 July, 2017\\
		Bexco, Busan, Korea}

\begin{document}

\section{Introduction}
The \hess \ array consists of four 12~m diameter Cherenkov Telescopes CT1--4,
built between 2002 and 2004 and a fifth, 28-m diameter telescope (CT5) built in
2012. Thanks to its large mirror surface, the minimum gamma-ray energy that CT5
can trigger on is $\sim 30\gev$. It operates at an event rate about ten times higher than that of CT1--4.  

The primary goal of the upgrade of the  CT1--4 cameras \cite{oldcameras} is
reducing the dead time of the cameras when they are operated together with CT5. The
dead time of the original CT1--4 cameras was $\sim 450 \eh{\mu s}$ which was
acceptable for the original array rates ($200-300\eh{Hz}$), but CT5 triggers
at $1.5\eh{kHz}$ or more, leading to a substantial fraction of events losing
CT1--4 information because of the dead time. The upgraded cameras are therefore
built around the \nectar \ analog memory chip \cite{nectar}, which allows for a
much lower dead time.

Another main goal of the upgrade of the CT1--4 cameras is reducing the failures
due to the aging of the electronics, which have been operated daily for 14
years in harsh conditions. 

The installation of the new cameras was completed by the end of October 2016.  
During the following months, the system was under commissioning and then in calibration and fine-tuning of the operation parameters. Around February 2017, the full \hessII\ array was working well and the new cameras were integrated with the data acquisition and CT5. Now
the array is  fully operational.

The following will report about the performance of the new components. The new hardware and
software is described in more detail in a second contribution to this conference \cite{hess1u_hwsw}.

\section{NECTAr calibration}
The analogue part of the readout is built around the \nectar ~analog memory chip \cite{nectar}.
The readout was developed to have very low noise. The
pedestal noise, measured as the RMS of the value of a single cell is around 4
ADC counts, or $2\eh{mV}$. At the nominal PMT gain of $3.2\times10^5$, this
corresponds to 0.2 photoelectrons.

\begin{figure}
  \centering
  \begin{minipage}[c]{0.61\textwidth}
  \includegraphics[width=1.\textwidth]{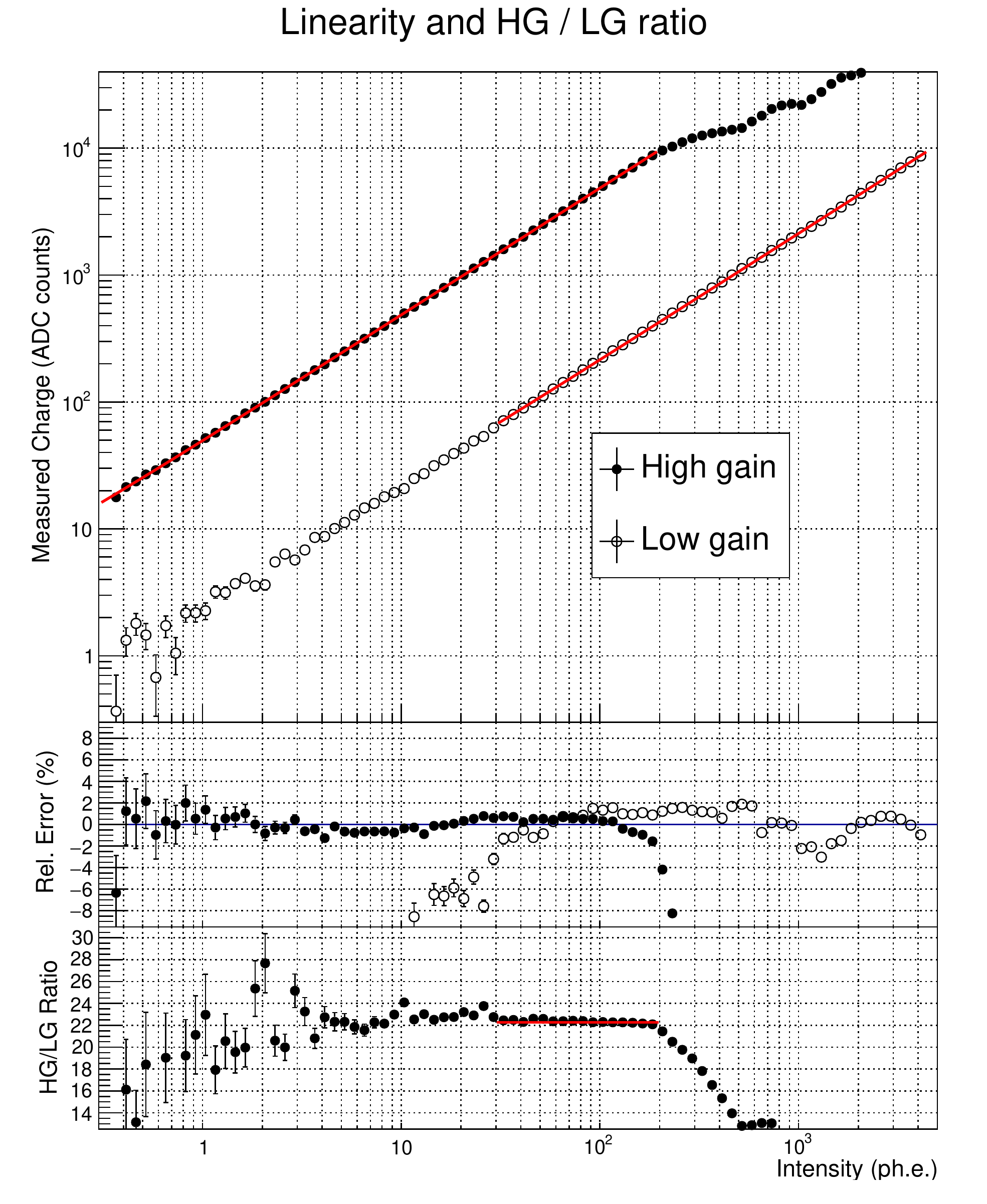}
  \end{minipage}
  \begin{minipage}[c]{0.37\textwidth}
  \includegraphics[height=.24\textheight]{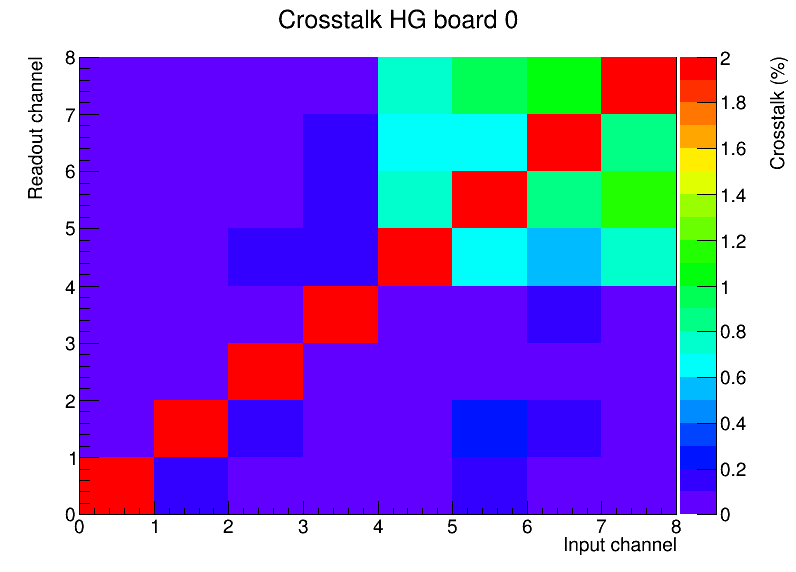} \\
  \includegraphics[height=.24\textheight]{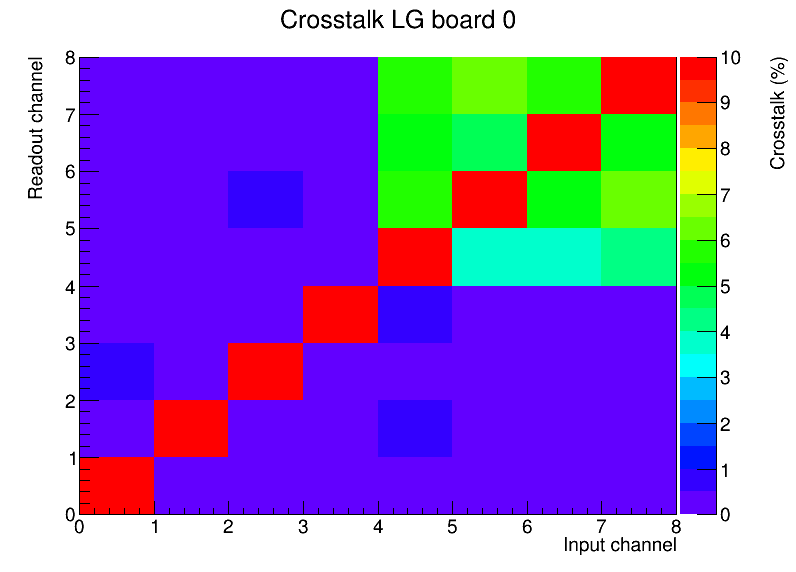}
  \end{minipage}
  \caption{Left: Summary of the linearity results for a typical channel. In the
  top frame, response curves are shown of both high gain (black filled circles)
  and low gain (empty circles) readout channels.   Right: Relative cross talk at the maximum of the linear range (about 200
  p.e. for HG, 4125 p.e. for LG). The higher cross-talk in the channels 4-7
  is due to the layout of the board. It is in any case below 1\% for the low gain branch and 5\% for the high gain branch.}
  \label{fig:linearity}
\end{figure}

The linearity of the high gain and low gain channels are on average better than
2\pcnt; The cross talk between two channels within one \nectar\ and
between channels on different \nectar s is typically less than 0.5\pcnt, and, depending on signal strenght,
never larger than 5\pcnt\ (see \fref{fig:linearity}, right).
The higher cross-talk in the channels 4-7 is due to the routing layout of the analog board.

The high gain linear range is between 0 and 200 photoelectrons (p.e.), and the low
gain one is between 30 and 4200 p.e.; the overall dynamic range of
this readout is therefore greater than 80~dB. The high to low gain ratio
remains constant at around 22 between 30 and 200 p.e. (see
\fref{fig:linearity}, left).

\begin{figure}
  \centering
  \includegraphics[scale=0.35]{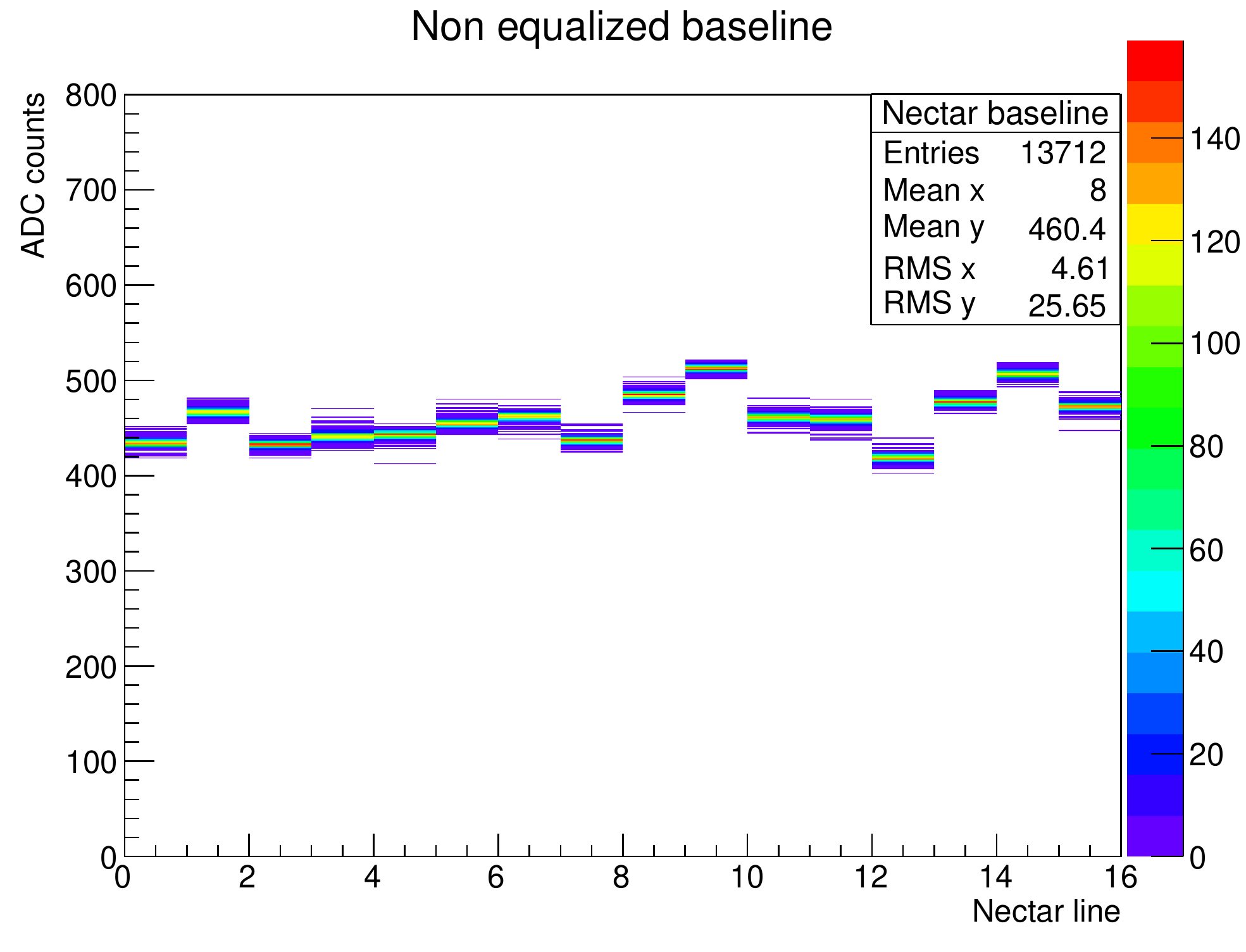}
  \includegraphics[scale=0.35]{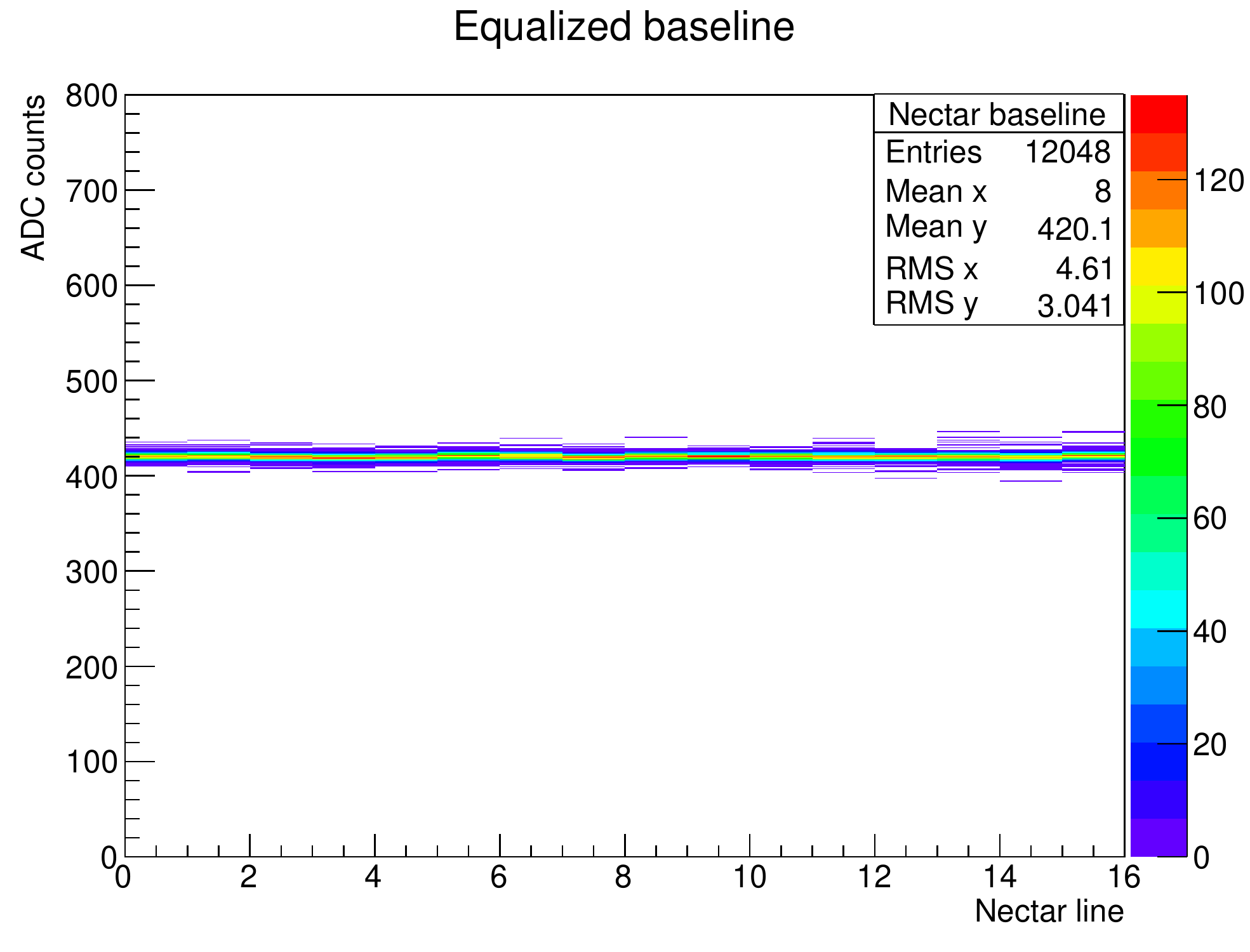} 
  \caption{Calibration of the \nectar ~chip baseline. Left: \nectar ~baseline before calibration. The mean value is higher than that expected and the RMS is $\sim$ 26 ADC counts. Right: \nectar ~baseline after calibration. The mean value is around 420 ADC counts as expected and the RMS is $\sim$ 3 ADC counts.}
  \label{fig:baseline}
\end{figure}

The \nectar\ inputs have a range of 2V \cite{nectardatasheet}. An adjustable constant common-mode offset is added
to the electrical signals just before it enters the chip, in order to account
for a possible undershoot and drifts in the baseline due to temperature
effects. Its default value is about $205\eh{mV}$ (420 ADC counts), but it is automatically
calibrated at the beginning of observation in order to have the same baseline for every
cell within the chip. \fref{fig:baseline} shows a comparison of the \nectar ~chip readout before and after baseline calibration.  The RMS of the baseline is lowered from  $\sim$ 20  down to $\sim$ 3 ADC counts after baseline calibration.

\section{Pixel calibration}
\begin{figure}
  \centering
  \includegraphics[height=.20\textheight]{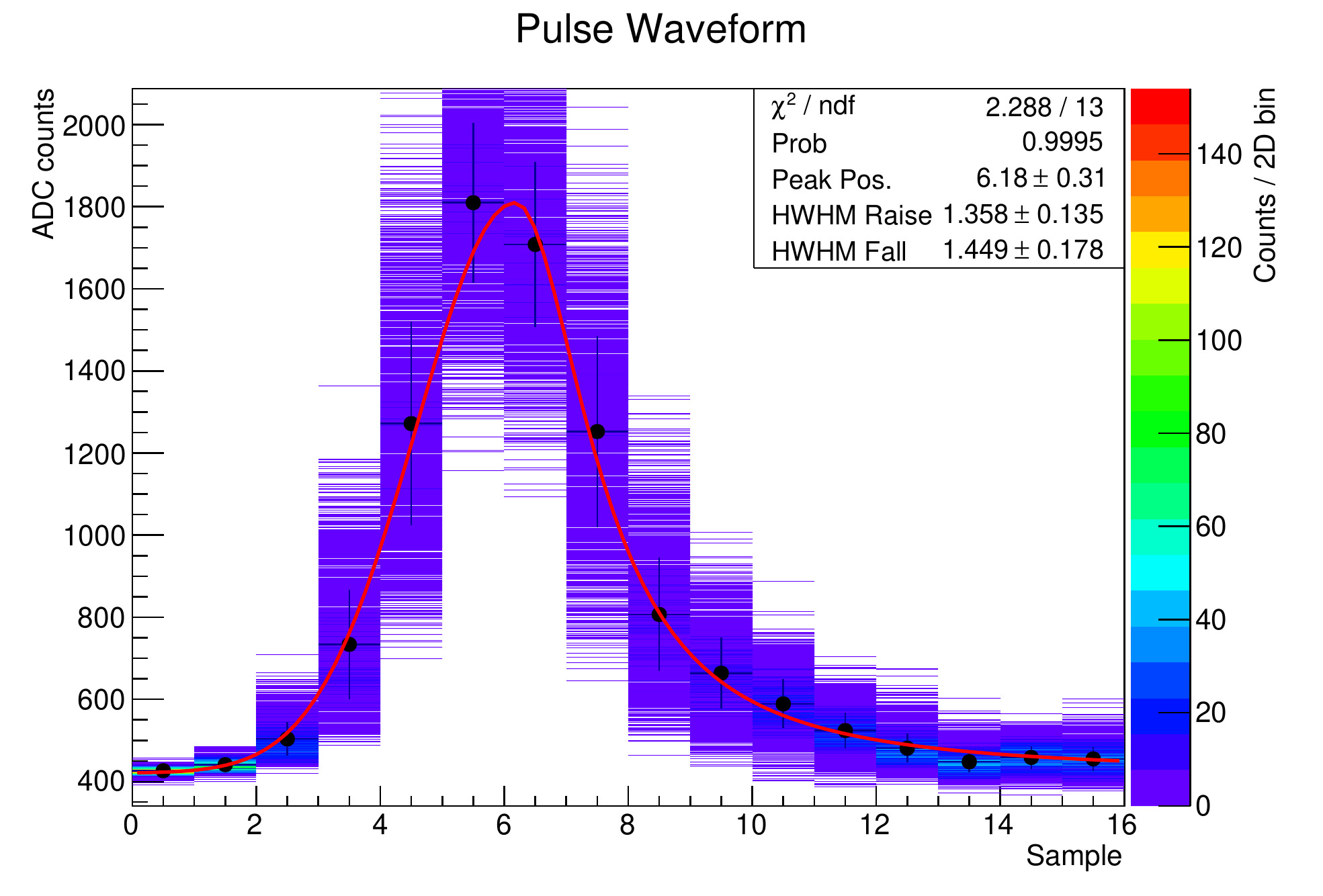}
  \includegraphics[height=.20\textheight]{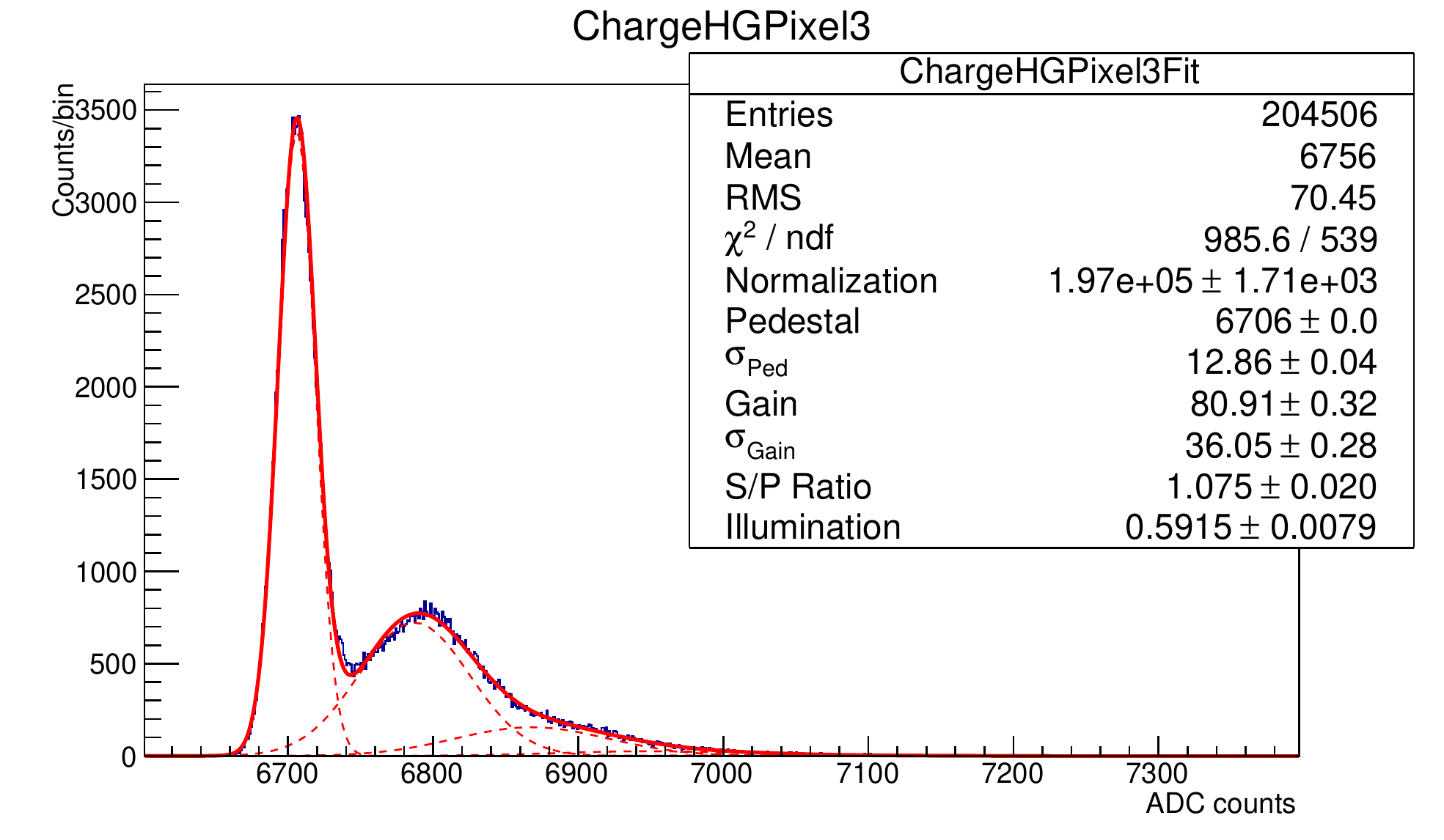}
  \caption{Left: Typical PMT pulse from a calibration source, as it is recorded by the
  readout.  Right: Results of a PMT gain calibration. The first peak is
  the pedestal peak, fitted to a gaussian function, the following ones are the one-, two-, etc. photoelectron peaks, also fitted to gaussian functions (the real single photoelectron charge distribution is not a gaussian, but this is corrected by applying a factor later in the analysis). The intrinsic electronic noise is the width of the pedestal peak, which for this PMT is about 12.9 ADC counts or 0.16 photoelectrons. The PMT gain is determined from the fit as the distance between the position (in ADC counts) of the pedestal peak and of the first photoelectron peak, and between each pair of the subsequent peaks.}
  \label{fig:gain_calib}
\end{figure}

In order to calibrate the gain of each PMT, a so-called single photo-electron (SPE) unit is used. The SPE unit consists in a light emitter placed in the shelters facing the cameras. 
The SPE unit can emit light pulsed at a wavelength of 370 nm with a frequency ranging from  $156\eh{kHz}$ down to $38\eh{Hz}$. The intensity of the pulses ranges from 1 to $200\eh{p.e.}$.
A diffuser is placed in front on the light emitter for a homogeneous illumination of the camera.\\
First, dedicated runs to calibrate the position of the readout window of the \nectar ~chip are performed: this is done by setting the SPE unit at high light intensity and adjusting the position of the readout window so that the PMT pulse is fully containted. The width of the readout window is fixed at 16 ns.\\
Then, a series of runs with varying PMT high voltage are taken, while the camera is illuminated by the SPE unit with a low light intensity. The gain of each PMT can be measured for each pixel and voltage setting from a fit of the charge distribution, as seen in \fref{fig:gain_calib}. The nominal high voltage of the PMT is reached when the nominal value of the gain is 80 ADC counts, or $3.2\times10^5$. This particular value is chosen so that the second peak, due to single photoelectron pulses, can easily be distinguished from the first, due to pedestal noise.
The resulting nominal high voltage of the PMTs ranges from $\sim800\eh{V}$ up to $\sim1200\eh{V}$. The above process is repeated several times to increase the accuracy of the gain flat-fielding.

\begin{figure}
  \centering
  \includegraphics[height=.2\textheight]{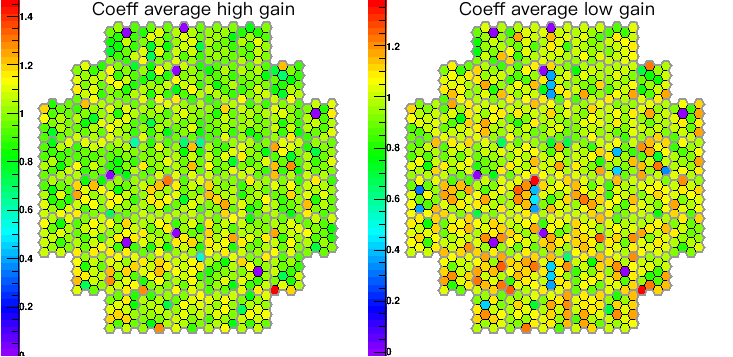}
  \includegraphics[height=.21\textheight]{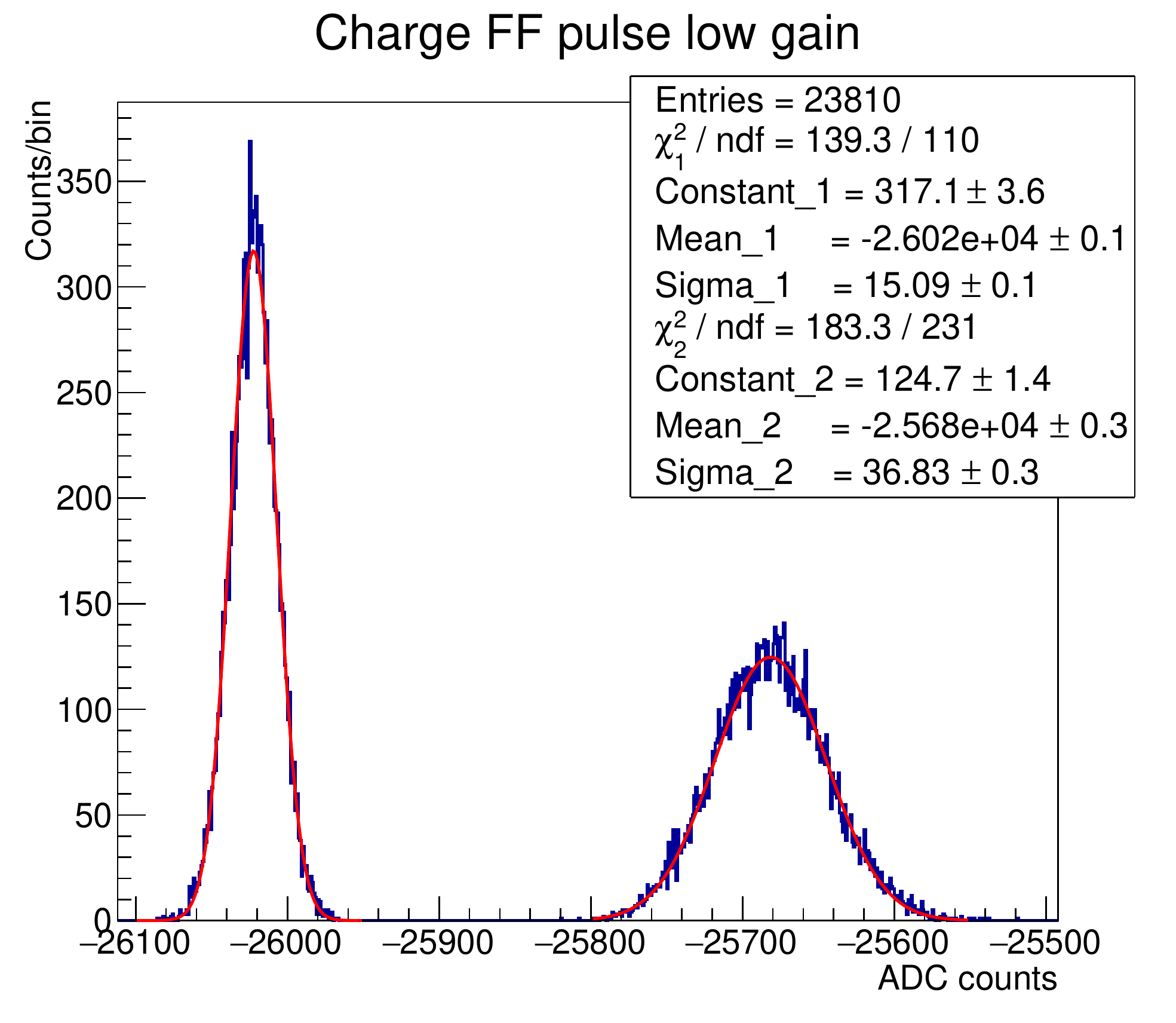}
  \caption{Response of the camera to light pulses from the flatfielding calibration unit.
  The calculation of the PMTs coefficients are shown for the whole camera using the high- (left) and low- (middle) gain branches. Right: Result of a PMT flatfield calibration. The first peak is the pedestal peak and the second peak is the distribution arising from the flatflied calibration unit.}
  \label{fig:ff_calib}
\end{figure}

In order to calibrate the different photo-cathode and funnel collection efficiencies for each pixel, a different device, called flat-fielding unit is used.
The flat-fielding units are located at the center of the mirrors of the telescopes and can emit short light pulses of moderate intensity ($\sim100$ p.e.) with a wavelength between 390 and 420 nm, around the PMT quantum efficiency peak.
A diffuser is mounted in front of the the light emitter to ensure a homogeneous illumination of the camera. The distribution of the flat-fielding coefficient across the camera for high- and low-gain, and the response in charge of a PMT during such a measurement are shown in \fref{fig:ff_calib}. The low gain charge distribution of a pixel shown there is fit to two Gaussian functions, one for the pedestal and one for the actual light pulses. The intensity of the light illumination, and therefore the correction coefficient are determined by the distance from the two peaks, which in this case corresponds to $\sim 94$ p.e.

The flatfielding unit is also used to determine the position of the readout window during regular observation runs, with a process similar to that of the SPE readout window position described previously. The typical flat-fielding pulse waveform is seen in Fig. \ref{fig:gain_calib}, left.

\section{Array trigger delay calibration}
The central array trigger is a key component of the stereoscopic observations. Its role is to collect the trigger signal from each telescope and search for coincident triggers in a given time windows set to 80 ns. The central trigger is located in the main building where also the servers are located and from where the telescopes are operated. Since each telescope is located at a different distances from the building, delays have to be taken into account when the triggers from several telescopes arrive to the central trigger. In order to measure the delay, a signal is sent via optical fiber from the central trigger to the cameras drawer interface box (DIB). The signal is then reflected and sent back to the central trigger where the time of flight is measured. The estimated delays were on average reduced by $\sim 300$ ns with respect to the previous system.

\section{Trigger threshold calibration}
The trigger system of \hess ~is organized in three layers. 
The lowest trigger layer (L0) is at the pixel level: the signal from each PMT is amplified and routed to a high-speed comparator, which issues an LVDS signal as long as the PMT signal is above a certain pixel threshold $q$.
The L0 signals in a half drawer (8 pixels) are summed together, converted to single-ended and routed to a board which provides the camera-wide (L1) trigger. This trigger is implemented by subdividing the camera into 38 overlapping sectors of 64 pixels each (8 half-drawers). For each sector, the signal from the half drawers are summed together and compared to a second threshold, $p$\footnote{Both $q$ and $p$ are voltage values. Their conversion into photoelectrons and pixels respectively requires dedicated calibration runs too, which are not explained here}. A L1 trigger is therefore issued whenever at least $p$ pixels in a camera sector are triggering simultaneously (majority logic). $p$ is set to 3 pixels by default. The L1 trigger signals are sent via a fiber-optic link to the central trigger, which is located in the control building of the array. 
There, the last layer of the trigger (array trigger) is implemented. It is a simple coincidence: in order for an event to be triggering the array (and thus the readout of the \hessI ~telescopes involved in it), at least two telescopes should issue a L1 trigger within a 80 ns wide coincidence window. The larger CT5 telescope however also records all monoscopic (CT5 L1 only) events.

\begin{figure}
  \centering
  \includegraphics[height=.6\textheight]{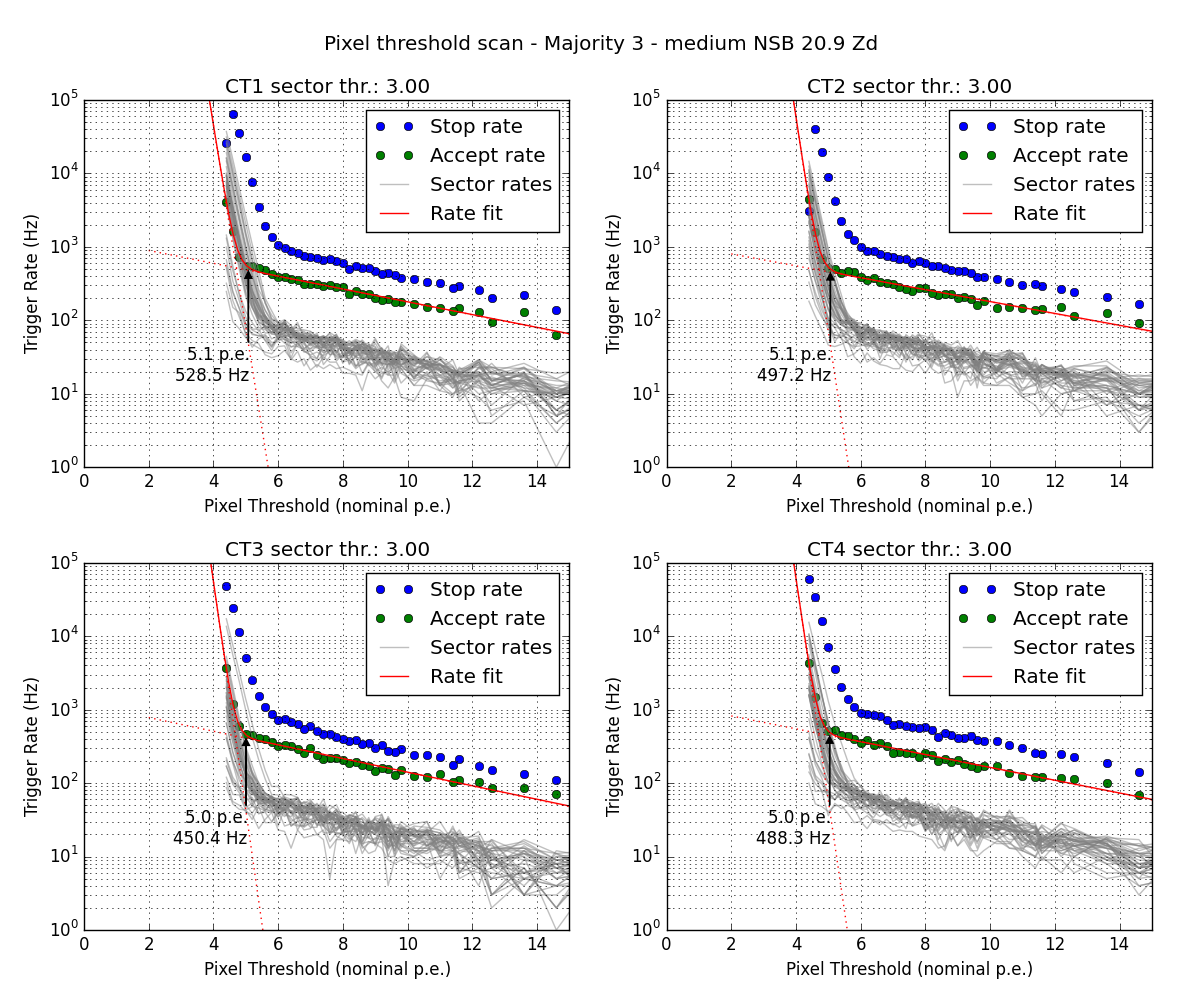}
  \caption{Results of the threshold array scan for the upgraded \hessI ~ cameras. The noise wall can easily be distinguished for thresholds lower than $\sim5$ p.e.. The blue and green points represent the L1 and stereo rates, respectively. The grey lines represent the sector rate for each of the 38 sectors of the cameras. The dashed red lines represent the fits to the noise wall and the cosmic-ray plateau, and the solid red line is the combined fit of both distributions.}
  \label{fig:ff_pixscan}
\end{figure}

The trigger rates are sensitive to observations conditions and position in the sky. The value of $q$ has to be calibrated in such a way that as many cosmic-ray air showers are detected, rejecting, at the same time, as much night sky background (NSB) as possible. This calibration is done by varying the value of $q$ and measuring the trigger rates. Such threshold scans are performed under optimal weather conditions and at low/medium zenith angles including the full array with CT5.
The results of the latest scans taken in March 2017 are shown in \fref{fig:ff_pixscan} for a Galactic source with a medium level of NSB. The figure depicts the sector, L1 and stereo trigger rates for each \hessI~telescope. In it, one can appreciate that the rates increase as $q$ decreases; below a certain value of $q$ however the trigger rates increase much faster. This is due to the system triggering on noise rather than on cosmic-ray showers (noise wall). In order not to have strong fluctuations in the recording rates, the amount of noise contamination at the the array trigger level (i.e. accidental triggers) was chosen to be 10\%. The value of the pixel threshold $q$ that fulfills this requirement is roughly 5 p.e., resulting in an array trigger rate of $\sim490$ Hz. 

\section{Sample mode}
When an event is triggered, the part of the PMT signal waveform sampled at 1 GHz which is inside the readout window is integrated, giving the ``charge'' deposited in each individual pixel. This was the only information sent to the DAQ in the previous system. 
Thanks to the advanced data acquisition electronic of the new cameras, additional information is now available to the analysers. In fact, the system is now able to send the position of the maximum sample inside the readout window (called time of maximum, ToM) and the total number of samples when the signal was above a certain threshold (time over threshold, ToT).
\\
In addition to charge, ToM and ToT, which are the standard data products, the readout can be configured so that the complete waveform information of up to 40 slices around the readout window is also stored. This is particluarly beneficial for inclined showers with energies higher than 1 TeV, where the Cherenkov light pulse recorded by the PMTs can last longer than $16\eh{ns}$. It is required to enhance the sensitivity of the system at the highest energies.
This ``sample'' datataking mode is now running on the \hessI ~cameras, with a readout window of 40 ns, operating at the standard nominal rate, in parallel to the standard mode.

\section{First observations}
\begin{figure}
  \centering
  \includegraphics[height=.4\textheight]{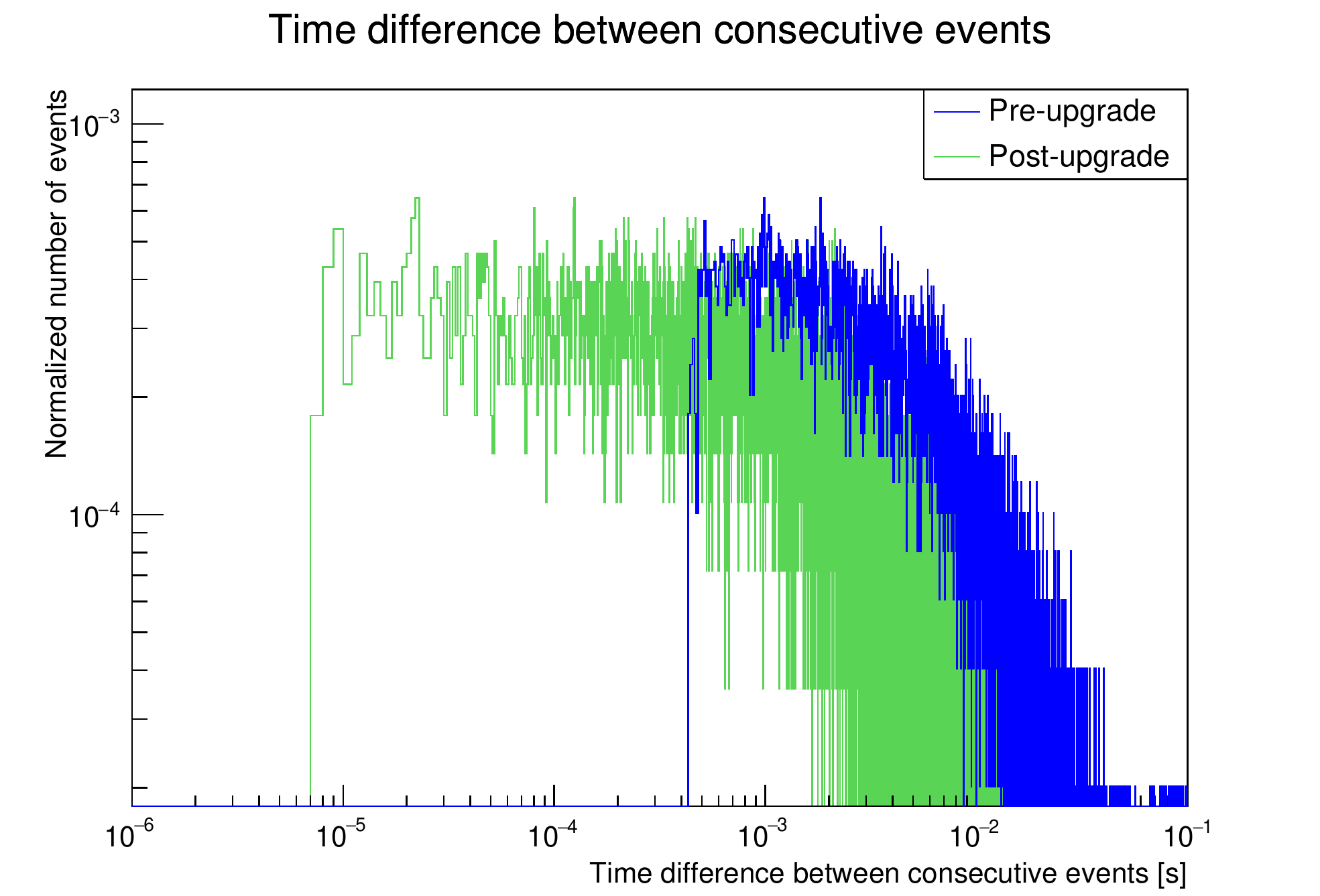}
  \caption{Comparison of the dead time before and after the upgrade. The blue and green distributions represent the normalized distribution of the time between consecutive events before and after upgrade, respectively. After upgrade, the minimum time between two events is $\sim7~\mu$s whereas it was $\sim400~\mu$s before the upgrade.}
  \label{fig:deadtime}
\end{figure}

The new cameras and their data acquisition and software was under commissioning until around February 2017. Meanwhile, observations were ongoing together with the calibration of the system and the first estimations of the performance of the new array. One of the main goals of the upgrade was to reduce the dead time after the trigger on a Cherenkov shower. The reduction of the dead time was tested comparing observations of Eta Carinae before and after upgrade. The result is shown in \fref{fig:deadtime}, using stereoscopic data from CT2. It can be seen that the smallest consecutive time between two events has been reduced from $\sim$\SI{400}{\micro\second} before upgrade, down to 
$\sim $\SI{7}{\micro\second} after upgrade. This result slightly differs from the first estimations of the dead time reported from the first camera in 2016 since now the delays from central trigger signal path are included in the measurement.

\begin{figure}
  \centering
  \includegraphics[height=.4\textheight]{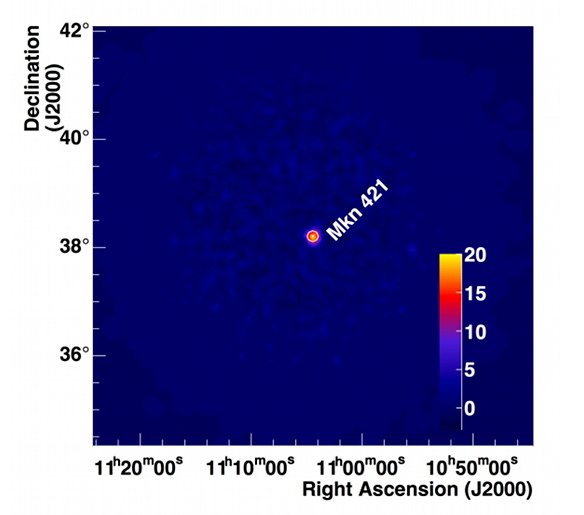}
  \caption{Sky map of the blazar Mkn 421, observed during the commissioning of the new 
  \hess ~cameras. Figure adopted from \cite{SOM_Mkn421}}
  \label{fig:SOM_Mkn421}
\end{figure}

At the beginning of January 2017, the blazar Mkn 421 was reported as being in a flaring state, by the HAWC collaboration\footnote{ATel \#9946}. The flare was observed by \hess ~ using the new system, collecting $\sim2$ hours of data. The data were processed using two independent analysis pipelines and preliminary sets of cuts. Both analysis revealed a clear detection at a 16$\sigma$ level of Mkn 421, being the first detection of a very-high-energy gamma-ray source using the \nectar ~technology \cite{SOM_Mkn421}. 
The resulting sky map is shown on \fref{fig:SOM_Mkn421}

\section{Summary}
The upgraded \hess ~I cameras were done installing at the end of October 2016 and fully commissioned and calibrated in February 2017. The first tests of the new system revealed a gain in the performance with an executive observation rate almost twice higher than that of the previous cameras. Furthermore, the  first observation during the commissioning period revealed a strong decrease in the dead time, which was one of the main purposes of the upgrade. Also, the detection of the blazar Mkn 421, at a 16 $\sigma$ level in $\sim$ 2 hours, during its flare state of January 2017, unvealed the efficiency of the \nectar ~chip that is to be used for the Cherenkov Telescope Array.~\newline

\end{document}